# Application of Multiple Imputation When Using Propensity Score Methods to Generalize Clinical Trials to Target Populations of Interest

Albee Y. Ling[1], Maria E. Montez-Rath[2], Kris Kapphahn[1], Manisha Desai[1]

[1]Quantitative Sciences Unit, Division of Biomedical Informatics Research, Department of Medicine, Stanford University School of Medicine, 1701 Page Mill Road, Palo Alto, CA 94304
[2]Division of Nephrology, Department of Medicine, Stanford University School of Medicine, 1070 Arastradero Road, Palo Alto, CA 94304

**Corresponding Author:**
Manisha Desai
Professor of Medicine and of Biomedical Data Science
1701 Page Mill Road, Room 237
Palo Alto, CA 94304
Quantitative Sciences Unit
Department of Medicine
Stanford University
(650) 725-1946
manishad@stanford.edu

**Abstract**

When the distribution of treatment effect modifiers differs between the trial sample and target population, inverse probability weighting (IPSW) can be applied to achieve an unbiased estimate of the population average treatment effect in the target population. The statistical validity of IPSW is threatened when there are missing data in the target population as well as in the trial sample. However, missing data methods have not been adequately discussed in the current literature. We conducted a set of simulation studies to determine how to apply multiple imputation (MI) in the context of IPSW. We specifically addressed questions such as which variables to include in the imputation model and whether they should come from trial or non-trial portion of the target population. Based on our findings, we recommend including in the imputation model as main effects all potential effect modifiers and trial indicator for both trial and non-trial population, as well as treatment and outcome variables from trial sample. Additionally, we have illustrated ideas by *transporting* findings from the Frequent Hemodialysis Network (FHN) Daily Trial to the United States prevalent hemodialysis population in 2017 defined using the United States Renal Data System.

# 1. Introduction

Randomized clinical trials (RCTs) are the gold standard when evaluating medical interventions. While RCTs have high internal validity, they suffer from low external validity.[1–4] The Frequent Hemodialysis Network (FHN) Daily Trial recruited 245 prevalent patients with end stage kidney disease (ESKD) during the study period of January 2006 to March 2010, with 120 individuals randomized to the conventional hemodialysis arm and 125 to the frequent hemodialysis arm. The trial found that frequent hemodialysis lowered risk of both coprimary outcomes [5]. The experimental protocol of more frequent hemodialysis, however, was not implemented in the clinical setting for several reasons including the fact that physicians remained uncertain as to whether the findings generalized to their patients. Thus, whether FHN trial findings can be applied to the "typical" patient on hemodialysis in the US remains unsolved. There are significant costs, both ethical and financial, associated with initiating new clinical trials for each target population of interest. Making use of existing data from trials can therefore be enormously valuable.

Recent work has led to promising methods to translate trial findings to target populations using propensity score (PS) based methods [2,3,6–12]. When a common set of treatment effect modifiers (variables that alter the effect of the exposure on the outcome) exist in both trial and target populations and their distributions differ in the two populations, the treatment effect estimate from a trial is not an unbiased estimate of that in the target population[2]. An approach known as inverse probability of sampling weighting (IPSW) can be applied here. Borrowing from the causal inference setting where the PS is defined as the conditional probability of treatment assignment, when translating findings, the PS describes the probability of membership in the trial sample conditional on pre-treatment covariates. Analogous to inverse probability of treatment weighting (IPTW), IPSW reweights the trial sample as a function of the PS, allowing the trial sample to resemble the target population[12]. This enables estimation of the population average treatment effect in the target population via weighted regression models[3,8,11,13]. By convention, we refer to research questions where the trial population is part of a broader target population of interest by *generalizability*; whereas, when the trial and target populations are disjoint, by *transportability*.[9,12,14,15]

While IPSW provides excellent solutions to translate trial findings in theory, validity may be threatened in practice. Specifically, the statistical validity of IPSW relies on a fundamental assumption that there are no unobserved variables associated with treatment effects and trial selection conditional on the observed covariates [3,16,17]. This assumption can be problematic in the presence of missing data. The issue of missing data involved in PS estimation is a crucial issue not adequately addressed in existing studies that translate clinical trial findings to target populations. The majority of the studies either did not comprehensively report the level of missingness in study variables or failed to mention any missing data methods, partly because critical variables with missing data were removed from further consideration in the analysis [2,3,7–9,18–23]. Some studies adopted complete cases analyses (CC) [24,25] or single imputation methods [16], which can lead to biased and inefficient estimates[26]. Studies that have applied multiple imputation (MI) methods [9,11,27–29] often lacked details on how PS was imputed or integrated in the presence of multiply imputed data sets. To our knowledge, there has only been one study that detailed its application of MI[30]. MI is a flexible tool for handling missing data with particularly appealing statistical properties[26], yet limited research has been done on how to use MI to handle missing data when applying IPSW in the context of translating trial findings to target populations and there is no consensus on the best statistical practice.

We propose an extensive simulation study to evaluate MI methods for handling missing data when IPSW is applied in the context of translating trial findings to target populations of interest. Applying MI to IPSW involves (1) specifying imputation models (2) imputing PS, and (3) integrating PS into the analysis to estimate the population average treatment effect. Based on previous research on (2) and (3) conducted in the context of PS matching and IPTW [31–33], we will adopt *MI-passive* (estimating PS after imputing the underlying missing variables) and *PSI-within* (the treatment effect is estimated by applying IPSW within each of the $m$ imputed data sets). In order to obtain a proper imputation model, all variables in the scientific model of interest, including treatment and outcome variables, are included[34,35]. However, the choice of imputation models in the context of *generalizability* and *transportability* is not as straightforward since treatment and outcome variables from the non-trial population are part of the scientific model. Thus, for (1), we will compare the performance of several imputation models, which differ on the underlying populations and variables included. In addition, we will vary the imputation model by inclusion and exclusion of interaction terms. Furthermore, we will illustrate principles learned in the simulation study by applying these tools to *transport* FHN results to the United States prevalent hemodialysis population in 2017 defined using the United States Renal Data System (USRDS) [36].

## 2. Methods

We have conducted a simulation study to assess the performance of various MI methods in handling missing data when IPSW is used to *generalize* trial findings to a target population.[11] Specifically, we assume that the trial is part of the target population, where PS is the probability of each individual being selected into the trial. Our simulations included scenarios where *generalization* of trial findings is particularly meaningful, i.e., where there is strong heterogeneity in the treatment effect and where the target and trial populations differ on key covariates or treatment effect modifiers. To that end, we generated a super-population to represent the target ($N_T$=10,000) comprised of both the trial sample ($N_t$=500) and non-trial population. There were 1,000 simulations generated, and the results were summarized across all simulations.

### 2.1 Data generation

*Covariates.* For the entire target population, we generated three covariates, $X = (X_1, X_2, X_3)$ from independent standard normal distribution.

*Trial indicator.* Within the target population, we defined the trial population through an indicator variable, $S$, where $S = 1$ and $S = 0$ indicated that a patient is in the trial or not, respectively. The probability of being in the trial was a function of all three covariates such that $logit(P(S = 1|X)) = \alpha X^T$ where $\alpha = (\alpha_0, 1,1,1)$ and $\alpha_0$ determines the relative sizes of the trial population. We set $\alpha_0 = -4.10$ for our main simulation scenario but varied its value in sensitivity scenarios (**Supplementary Table 1**).

*Random treatment assignment.* Among those in the trial ($S = 1$), treatment status, denoted as $A$, was generated from a Bernoulli distribution with probability 0.5.

*Potential outcomes.* Under the potential outcomes framework [37,38], we generated true outcomes under both treatment and control for all patients in the super-population. Let $Y^1$ and $Y^0$ denote the outcomes corresponding to a patient treated or not respectively:

$$Y^1 = \boldsymbol{\beta^1}^T(1, X) + \varepsilon^1 \; ; \; Y^0 = \boldsymbol{\beta^0}^T(1, X) + \varepsilon^0$$

where $\boldsymbol{\beta^1} = c(1,1,1,1)$ and $\boldsymbol{\beta^0} = c(0,0,0,1)$. By varying the coefficients when generating $Y^0$ and $Y^1$, we were able to simulate treatment effect modification.

*Observed outcome.* Among those in the trial, while everyone had two potential outcomes, only one was realized such that $Y = (1 - A)Y^0 + AY^1$.

*Sensitivity simulations.* Trial and target sizes were varied in sensitivity simulations, while all other aspects of the simulations were kept the same. Please see **Supplementary Table 1** for values of $\alpha_0$ in sensitivity simulation studies.

In the absence of missing data, we directly estimated the true population average treatment effect in the target population, $PATE = E[Y^1 - Y^0 | S = 0]$ using the outcomes in the entire target population. We also obtained the treatment effect estimated from the trial population, $TATE = E[Y | S = 1, A = 1] - E[Y | S = 1, A = 0]$.

## 2.2 Generalizability using IPSW

To apply IPSW, we estimated PS coefficients using a correctly specified logistic regression model by regressing $S$ on $X = (X_1, X_2, X_3)$. PSs were estimated as the fitted values of the logistic regression model on the probability scale. Weights were generated for all trial patients as the following:

$$w_i = \frac{S_i I(A_i = a)}{\hat{p}(X)\widehat{e_a}(X)}$$

$\hat{p}(X)$ is the estimated PS and $\widehat{e_a}(X)$ is an estimate of the conditional probability of treatment assignment in the trial sample ($\Pr[A = a | X, S = 1])$[11].
Using these weights, a weighted linear regression model was fitted in the trial sample only and $PATE_{ipsw}$ was estimated as the regression coefficient for the treatment variable[2,9,39].

## 2.3 Missing data mechanisms (MDMs)

We induced missingness in one of the treatment effect modifiers, $X_1$, and assumed all the other variables were fully observed. The MDMs considered were missing at random (MAR) scenarios, where the top 30% of subjects ranked by the sum of their $X_2, X_3$ in the non-trial population were missing. While there was always 30% missing data in the non-trial population, the proportion missing in the trial varied (0%, 10%, and 30%).

## 2.4 Missing data methods

We considered the following MI imputation models in our simulation study:

M1A (covariates-only-non-trial): $X_1 \sim X_2 + X_3$ in non-trial population only

M1B (covariates-only-superpopulation): $X_1 \sim X_2 + X_3$ in both trial and non-trial populations

M2 (all-variable-no-interaction): $X_1 \sim X_2 + X_3 + X_4 + S + Y + A$ in both trial and non-trial populations

M3A/B (all-variable-with-interaction): $X_1 \sim X_2 + X_3 + S + Y + A + X_1 * A + X_2 * A$ in both trial and non-trial populations. Under M3A, *active* imputation – i.e., imputing all terms including interaction terms as any other variable was performed for the interaction terms. In contrast, M3B relies on *passive* imputation, which involves imputing in a series of steps where dependencies of a derived variable that is a function of other variables are acknowledged. Note that under M2 and M3A/B, $A$ and $Y$ were observed only in the trial but not in the non-trial population. $A$ and $Y$ in the non-trial population were imputed to aid the imputation of missing variable, $X_1$, but did not play a role in the subsequent estimation of the treatment effect.

After specifying imputation models, *MI-passive PSI-within* was adopted for IPSW analysis[33]. Under *MI-passive PSI-within*, covariates were imputed first before PS was estimated and the complete IPSW analysis was conducted within each imputed dataset. When executing approaches in the R Package MICE, the number of multiply imputed datasets, $m$, was selected to be the same as the percentage of missing data in the entire target population [40]. For example, for the main simulation scenario, $m = 29, 29,$ and $39$, when the trial was missing data on $0\%$, $10\%$, and $30\%$ respectively. Five iterations ($maxit = 5$), and default settings for the imputation model (predictive mean matching for continuous variable and logistic regression for binary) were used[41]. As a comparator, we also considered complete case (CC) analysis, a commonly applied missing data method, where individuals with any missing data are excluded from the analysis.

## 2.5 Variance estimation

Empirical standard errors (SEs) were estimated as the standard deviation of $PATE$ estimates across 1,000 simulations.

## 2.6 Performance metrics

For each missing data method, we report on average bias, empirical SE, average robust SE, and average mean squared error (MSE), summarized over 1,000 simulations with estimated Monte Carlo SEs.[42]

## 3. Simulation Results

Under the main simulation scenario, we had 10,000 patients in the entire target population, 500 of whom were selected into the trial. The true average treatment effect in the target population, $PATE$, was estimated to be -1.001. The average treatment effect estimated in the trial population, $TATE$, was estimated to be -2.566 with empirical SE 0.150. When there were no missing data in trial or non-trial population, the IPSW analysis produced an estimate of $PATE$ with average bias -0.072, empirical SE 0.581, and average MSE 0.342 (Table 1).

Among all missing data methods, CC had the second largest bias and the largest empirical SE, where both bias and inefficiency contributed to the suboptimal MSE (Table 1 and Figure 1). Among the various MI methods, M1B yielded the largest bias and incorrectly underestimated the empirical variance, compared to the estimates using full data without missingness (Table 1 & Figure 1). M1A, which is only applicable when there was no missing data in the trial population, also underestimated the empirical variance despite its small bias. M2, M3A, and M3B achieved similar performance with respect to bias, variance, and mean squared error across all levels of missingness in the trial. Their performances were comparable to IPSW estimates when there was no missing data (Table 1 & Figure 1).

The comparative performance of various missing data methods was similar in sensitivity simulations (Supplementary Tables 2-8 & Supplementary Figures 1-7). In our main simulation scenario, the average bias, empirical SE, and average MSE produced using M2 were like the values using full data without missingness. However, this was not the case for Simulation Scenarios 5-8, where there was a sizable gap between the two sets of estimates. Interestingly, except for our main simulation scenario (Simulation Scenario 2) and Simulation Scenario 1, M1A performed well in estimating the empirical SE, in addition to its small bias.

## 4. Transporting FHN Trial Results to USRDS Patients

We illustrated the performance of various missing data methods in *transporting* FHN findings to a target population of all the United States prevalent patients on hemodialysis in 2017 defined using the USRDS.

### *4.1 Trial Population*

There were 245 patients in the original FHN trial, 120 undergoing conventional hemodialysis (three times a week) and 125 undergoing frequent hemodialysis (six times a week) (**Table 2**)[5]. The trial reported benefits in both co-primary outcomes: 1) death or change in left ventricular mass (LVM) (hazard ratio (HR) [95% confidence interval (CI)]: 0.61 [0.46 to 0.82]); and 2) death or change in the RAND-36 physical health composite (PHC) score (HR [95% CI]: 0.70 [0.53 to 0.92])[5].

### *4.2 Target Population*

The USRDS is a national registry of all patients receiving treatment for ESKD in the US and was used to derive the target population [36]. Baseline patient characteristics were largely derived from Medical Evidence Reports submitted by nephrologists within 45 days of a patient initiating ESKD treatment. These forms provided data on demographic characteristics, cause of ESKD, initial ESKD treatment modality, recent or current comorbid conditions, current laboratory measures, and current insurance payer at the time of ESKD treatment initiation, all of which may be key for estimating the PS. To *transport* the FHN results to the typical ESKD patient in the US, we defined our target population as all prevalent patients who became ESKD after Jan 1st, 1996 undergoing in-center hemodialysis on Jan 1st, 2017 (n=446,976). A comparison of patient characteristics between the FHN trial sample and the target population indicates that the trial included younger patients, fewer women, fewer white patients, and patients with longer ESKD duration than those captured in the USRDS (**Supplementary Table 9**).

### *4.3 Statistical methods*

Data from the trial sample and target population were concatenated and an indicator variable was created to indicate trial membership. Logistic regression was used to estimate PS, defined as the probability of being in the trial, from pre-treatment covariates: age, sex, race, cause of ESKD, and duration of ESKD (vintage). The weights were calculated based on the estimated PS as below[12]:

$$w_i = \begin{cases} \dfrac{P(S_i = 0 \mid X_i)}{P(S_i = 1 \mid X_i)} \times \dfrac{P(S_i = 1)}{P(S_i = 0)}, & S_i = 1 \\ 0, & S_i = 0 \end{cases}$$

Next, we fit a weighted Cox regression model of *Y* on *A* in the trial population only, using the weights estimated above. Bootstrap variance was reported, and 2,000 bootstrap samples were used. We implemented CC and MI (M1B, M2, M3A, and M3B) as missing data methods; M1A was not applicable since there were missing data in the trial as well as in the target population. For MI, m=5 imputed datasets were used since the overall missing level in the concatenated dataset was less than 2%[40]. Absolute standardized differences (ASD) were estimated before and after weighting in each imputed dataset to evaluate balance in baseline covariates [43,44]. Tipton index was also estimated in each of the imputed datasets to assess the similarity between trial and target populations as well as quantify the effectiveness of the propensity score adjustment [45].

*4.4 Results and Interpretation*

The Tipton index values calculated for both outcomes across all imputed datasets were between 0.79 and 0.80, falling in the medium to high range and indicating that a generalizability study is possible using these data but where we should expect some bias and loss of efficiency (**Table 2**) [45]. The treatment effects estimated using MI methods were more similar amongst themselves than to that of CC for both co-primary outcomes. The difference between MI methods was not as pronounced as that observed in the simulation studies possibly due to the low level of missingness. For the outcome of death or change in LVM, the *transported* HRs to the target population ranged from 0.64 to 0.67, which is slightly larger than the trial treatment effect estimate (0.61) (**Table 2**). For the outcome of death or change in PHC, the *transported* HRs were between 0.61 and 0.71, which includes the trial treatment effect estimate (0.70) (**Table 2**). Regardless of the missing data method, all the *transported* HRs fell into the 95% confidence interval of that estimated in the FHN trial for outcome LVM, rendering the trial and target populations in *estimate agreement* [46]. However, for PHC, *estimate agreement* was only achieved using MI (M1B, M2, and M3B). Note that the *transported* results are much less precise than the trial results, making it difficult to achieve statistical significance for both outcomes. This means that the trial and target treatment effect estimates in *regulatory disagreement* with each other [46]. Based on the *transported* results using M2 (which demonstrated superior statistical properties from our simulations), the beneficial effect of frequent hemodialysis was slightly larger in the target population for both outcomes. However, the confidence intervals were too wide to conclude that the FHN trial findings *transport* to our target population.

**5. Discussion**

To our knowledge, we are the first to use simulations to study the impact of missing data and the application of MI in the context of *generalizing* clinical trial results to the target populations of interest. We have simulated scenarios where strong effect modification existed so that the average treatment effects in the trial did not reflect that of the target population. Based on our simulation results and computing considerations, we recommend using the superpopulation and including the covariates, the trial indicator, treatment, and outcome, without interaction terms to impute missing data (M2), given its modest bias, empirical SE, and MSE.

Compared to M1A and M1B, M2 additionally included trial population in the imputation model, which more accurately reflected the scientific model for our IPSW analysis. Unlike M3A and M3B, however, M2 is not a proper imputation model[34,35,47–49] because it does not include the interaction terms. Nevertheless, M2 performed comparably to the imputation models that correctly reflect the true missing data mechanism. Further, it is logistically less difficult to apply, especially when the number of potential effect modifiers is large.

We did not recommend either M1A or M1B. In both imputation models, not all relationships specified in the scientific model are included, as required by a proper imputation model [34,35,47–49]. Even in the case of exception under Simulation Scenario 4, where M1A achieved excellent performance, M2 performed similarly. M1A is also limited because it can only be applied when there is no missing data in the trial population. Unfortunately, M1B did not perform well with respect to bias or variance estimation. Since trial indicator, treatment, and outcome were not part of the imputation model in M1B, there was less variation in the imputation process, resulting in a low variance. On the other hand, important interrelationships between variables were left out of the imputation process, leading to large bias.

There are limitations to our study. Firstly, we did not explore how to best estimate variance when MI was used in the context of IPSW. If using robust SE, our recommended MI imputation model, M2, would underestimate empirical SE (Table 1 & Supplementary Tables 2-8). Other variance estimators such as bootstrap methods are a more suitable alternative[11]. Secondly, we only briefly explored the impact of sample size and/or the trial to target ratio on the performance of MI. In our simulations with large target population size such as 100,000 in Simulation Scenarios 5-8, even the best performing MI method could not recover the average treatment effect estimated using full data. One hypothesis is that it is increasingly difficult to achieve desirable statistical properties with *any* missing data methods when the target population is too large or the trial to target ratio is too small. Future studies will need to be conducted to test this hypothesis and explain our simulation findings. Thirdly, our study exclusively focused on imputing missing data from trial and target populations in one concatenated dataset, so the imputation model can be congenial with the scientific model [35]. Mollan et. al chose to conduct MI separately in trial and target populations [30], possibly because data for these two populations are usually collected separately and their MDMs may differ as a result. Future researchers could compare these two ways of conducting MI to assess if their relative performance differs based on the similarity of trial sample and target population. For example, the trial sample and target population can come from two vastly different data resources in *transportability* scenarios, in which case imputing separately may yield better performance.

Other limitations include that there were only three independent covariates in our simulations, all of which are associated with both trial selection and outcomes. In practice, there would be many more variables, some of which are related to trial indicator or outcomes only. Nevertheless, this simulation setup was sufficient to demonstrate that certain models should not be used when conducting these studies and would be expected to perform worse had we included more complexity. We also only simulated linear relationships in selection and outcome generating functions, while the performance of MI in the context of non-linear relationships need to be evaluated as well. Missing data was induced in only one effect modifier in the target population, whereas more variables are likely to be missing in both trial and target populations in the real world. Additionally, since our primary focus was on missing data methods, we did not study variable selection into PS model or the effect of PS model misspecification.

To conclude, we have demonstrated that MI should be used when there are missing data in potential effect modifiers when IPSW is applied to *generalize* or *transport* trial findings to target populations of interest. Specially, the MI imputation model should include all covariates and trial indicator in both trial and non-trial populations, and treatment and outcome variables in the trial population. This imputation model coupled with *MI-passive* (impute PS variables and then estimate PS) and *PSI-within* (conduct IPSW analysis within each imputed dataset) [33] will yield results with the best statistical properties.

## 6. Acknowledgements

This work was supported by a Sanofi iDEA Award. We thank Glenn Chertow and Tom Greene for their excellent guidance and for sharing their knowledge of the FHN. The Frequent Hemodialysis Network Daily Trial (FHN Daily) was conducted by the Frequent Hemodialysis Network Daily Trial Investigators and supported by the National Institute of Diabetes and Digestive and Kidney Diseases (NIDDK). The data from the Frequent Hemodialysis Network

**Table 1.** Performance of different missing data methods when data is missing at random (MAR) in the main simulation scenario (nTrial = 500, nTarget = 10,000). Mean (standard deviation) of 1,000 simulation results are reported here. SE = Standard error; N/A= Not available; CC = Complete Case Analysis; MSE = mean squared error; M1A-M3B indicate various imputation models when multiple imputation was applied (see Methods). Using trial data only, treatment effect was estimated to be -2.57 (SE=0.15).

| Missing Data Method | Trial Missing Level | Bias | Empirical SE | MSE |
|---|---|---|---|---|
| Full Data | N/A | -0.072 (0.018) | 0.581 (0.013) | 0.342 (0.023) |
| CC | 0 | 0.471 (0.042) | 1.341 (0.030) | 2.018 (0.097) |
| CC | 0.1 | 0.474 (0.042) | 1.339 (0.030) | 2.015 (0.097) |
| CC | 0.3 | 0.478 (0.042) | 1.331 (0.030) | 1.998 (0.096) |
| M1A | 0 | 0.000 (0.010) | 0.300 (0.007) | 0.088 (0.003) |
| M1B | 0 | -0.624 (0.008) | 0.260 (0.006) | 0.456 (0.010) |
| M1B | 0.1 | -0.603 (0.008) | 0.268 (0.006) | 0.434 (0.010) |
| M1B | 0.3 | -0.600 (0.009) | 0.287 (0.006) | 0.440 (0.011) |
| M2 | 0 | -0.077 (0.018) | 0.568 (0.013) | 0.329 (0.017) |
| M2 | 0.1 | -0.066 (0.017) | 0.542 (0.012) | 0.297 (0.013) |
| M2 | 0.3 | -0.030 (0.017) | 0.552 (0.012) | 0.304 (0.016) |
| M3A | 0 | -0.062 (0.018) | 0.581 (0.013) | 0.341 (0.024) |
| M3A | 0.1 | -0.027 (0.019) | 0.607 (0.014) | 0.369 (0.027) |
| M3A | 0.3 | -0.017 (0.018) | 0.572 (0.013) | 0.327 (0.022) |
| M3B | 0 | -0.064 (0.018) | 0.580 (0.013) | 0.340 (0.023) |
| M3B | 0.1 | -0.066 (0.018) | 0.575 (0.013) | 0.334 (0.026) |
| M3B | 0.3 | -0.017 (0.018) | 0.575 (0.013) | 0.330 (0.023) |

**Table 2.** Estimated hazard ratio (95% confidence interval) in the FHN trial and transported to the target population derived from the United States Renal Data System by missing data method. LVM = death or change in left ventricular mass; PHC = death or change in the RAND-36 physical health composite score. CC = Complete Case Analysis; M1A-M3B indicate various imputation models when multiple imputation was applied (see Methods).

|  | LVM | PHC |
|---|---|---|
| **FHN Trial** | 0.61 (0.45, 0.82) | 0.70 (0.53, 0.92) |
| **CC** | 0.67 (0.57, 1.24) | 0.61 (0.53, 1.22) |
| **M1B** | 0.65 (0.54, 1.28) | 0.70 (0.57, 1.29) |
| **M2** | 0.64 (0.54, 1.27) | 0.71 (0.57, 1.30) |
| **M3A** | 0.64 (0.54, 1.27) | 0.69 (0.56, 1.28) |
| **M3B** | 0.64 (0.54, 1.27) | 0.70 (0.57, 1.29) |